\documentclass[proceedings]{JHEP3} 

\conference{International Europhysics Conference on HEP}

\usepackage{epsfig}                   

\title{Beyond ``naive'' factorization\\
 in  exclusive radiative $B$\/--meson decays}

\author{\speaker{Thorsten Feldmann}\thanks{Based on work together with
    M.~Beneke and D.~Seidel \cite{Beneke:2001at}.}\\
        Institut f\"ur Theoretische Physik~E, RWTH Aachen, 52056 Aachen, Germany \\
        E-mail: \email{feldmann@physik.rwth-aachen.de}}

\abstract{We apply the QCD factorization approach to exclusive, radiative 
 $B$ meson decays in the region of small invariant photon mass. 
 We calculate factorizable and non-factorizable
 corrections to leading order in the heavy quark mass expansion and
 next-to-leading order in the strong coupling constant. 
 Phenomenological 
 consequences for the $B \to K^*\gamma$ decay rate and the $B \to
 K^*\ell^+\ell^-$ forward-backward
 asymmetry are discussed.}

\newbox\mybox
\newcommand\fverb{\setbox\mybox=\hbox\bgroup\verb}
\newcommand\fverbdo{\egroup\medskip\noindent\fbox{\unhbox\mybox}\ }
\newcommand\fverbit{\egroup\item[\fbox{\unhbox\mybox}]}

\begin{document}

Radiative $B$\/--meson decays provide an important tool to test
the standard model of electroweak interactions and to
constrain various models of new physics.
The theoretical description of {\em exclusive}\/ channels 
has to deal with hadronic uncertainties related to the binding of quarks
in the initial and final states. 
For the decays $B \to K^*\gamma$ and $B\to K^*\ell^+\ell^-$, that we are
focusing on here, this is usually phrased as the need to know the hadronic form 
factors for the $B\to K^{(*)}$ transition, 
but there also exist ``non-factorizable'' strong interaction effects 
that do not correspond to form factors. They arise from 
the matrix elements of purely hadronic operators in 
the weak effective Hamiltonian with a photon radiated from one of
the internal quarks. In Ref.~\cite{Beneke:2001at}
we have computed these non-factorizable 
corrections and demonstrated that exclusive, radiative 
decays can be treated in a similarly systematic manner as their inclusive 
counterparts. As a result we obtain the branching fractions for 
$B\to K^*\gamma$ and $B\to K^*\ell^+\ell^-$ for small invariant mass 
of the lepton pair to next-to-leading logarithmic (NLL) order in 
renormalization-group improved perturbation theory.

In the ``naive'' factorization approach, exclusive radiative
$B$ decays are described in terms of 
hadronic matrix elements of the electromagnetic penguin operator
${\cal O}_7$ and the semi-leptonic
operators ${\cal O}_{9,10}$ \cite{Buchalla:1996vs}.
These are parametrized in terms of the corresponding tensor, vector
and axial-vector $B \to K^*$ transition form factors 
($T_{1,2,3}(q^2)$, $V(q^2)$, $A_{0,1,2}(q^2)$).
Factorizable quark--loop contributions (Fig.~\ref{fig1}b)
with the four--quark operators
${\cal O}_{1-6}$ are taken into account by using ``effective'' 
Wilson-coefficients, $C_7 \to C_7^{\rm eff}$, $C_9 \to C_9^{\rm eff}(q^2)$,
renormalized at the scale $\mu = m_b$.

\FIGURE[ht]{
\psfig{file=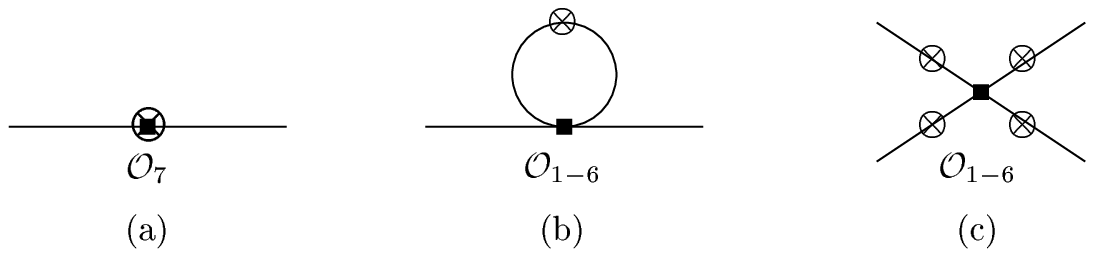, width=9cm}
\caption{LO contributions to 
$\langle \gamma^*\bar K^*| H_{\rm eff} | \bar{B}\rangle$. 
The circled cross marks the possible insertions of the virtual 
photon line. In (a) and (b) the spectator line is not shown.
}
\label{fig1}
}

In order to include non-factorizable contributions as in
Fig.~\ref{fig1}c and Fig.~\ref{fig2} it is convenient to introduce 
generalized form factors ${\cal T}_i(q^2)$ for the transition into a
{\em virtual}\/ photon $B\to K^*\gamma^*$
as follows,
\begin{eqnarray}
&& \langle \gamma^*(q,\mu) \bar K^*(p',\varepsilon^*)| H_{\rm eff} | \bar{B}(p)
\rangle = -\frac{G_F}{\sqrt{2}}\,V_{ts}^* V_{tb}\,\frac{i g_{\rm em}
  m_b}{4\pi^2}\nonumber\\[0.0cm]
&& \qquad 
\Bigg\{2 \,{\cal T}_1(q^2) \,\epsilon^{\mu\nu\rho\sigma}
\varepsilon^{\ast}_\nu\, p_\rho p^{\prime}_\sigma
-i\,{\cal T}_2(q^2)\left[(M_B^2-m_{K^*}^2)\,\varepsilon^{\ast\mu}-
(\varepsilon^\ast\cdot
q)\,(p^\mu+p^{\prime\,\mu})\right]
\nonumber\\[0.0cm]
&& \qquad 
-i\,{\cal T}_3(q^2)\,(\varepsilon^\ast\cdot
q)\left[q^\mu-\frac{q^2}{M_B^2-m_{K^*}^2}(p^\mu+p^{\prime\,\mu})\right]
\Bigg\} \, .
\label{caltdef}
\end{eqnarray}
In the ``naive'' factorization
approach these new functions reduce to 
${\cal T}_i(q^2)= C_7^{\rm eff} \, T_i(q^2) + \ldots$
Following the QCD factorization approach to exclusive $B$ decays
\cite{Beneke:1999br},
factorizable and non-factorizable radiative corrections 
are calculable in the heavy quark mass limit 
and for small photon virtualities (in practice $q^2 < 4 m_c^2$).

At leading order (LO) in the strong coupling constant, the 
generalized form factors read
\begin{eqnarray}
  \label{firstT}
{\cal T}_1(q^2) &=& C_7^{\,\rm eff} \,T_1(q^2) + Y(q^2) \,\frac{q^2}
{2 m_b (M_B+m_{K^*})}\,V(q^2), \nonumber \\
{\cal T}_2(q^2) &=& C_7^{\,\rm eff} \,T_2(q^2) + Y(q^2) \,\frac{q^2}
{2 m_b (M_B-m_{K^*})}\,A_1(q^2), \nonumber\\
{\cal T}_3(q^2) &=& C_7^{\,\rm eff} \,T_3(q^2) + Y(q^2) \,\left[\frac{M_B-m_{K^*}}
{2 m_b} \,A_2
(q^2)- \frac{M_B+m_{K^*}}{2 m_b}\,A_1(q^2)\right]
\cr && 
- e_q \, (C_3 + 3 C_4) \,
     \frac{8 \pi^2 M_B f_B f_{K^*} m_{K^*}}{N_C m_b (M^2-q^2)}  
      \int d\omega \frac{\phi_{B,-}(\omega)}{\omega -
         q^2/M - i \epsilon}. 
\label{lastT}
\end{eqnarray}
The function $Y(q^2)$, which is usually absorbed into
$C_9^{\rm eff}(q^2)$, arises from the quark loop in
Fig.~\ref{fig1}b. The last, ``non-factorizable'' 
term in ${\cal T}_3(q^2)$ comes from the
annihilation graph in Fig.~\ref{fig1}c when the photon is emitted
from the light spectator in the $B$ meson (all other graphs are
sub-leading in the $1/m_b$ expansion). It introduces a new
non-perturbative ingredient, namely one of the
two light-cone distribution amplitudes of the $B$ meson,
$\phi_{B,\pm}(\omega)$, see \cite{Beneke:2001at,Beneke:2001wa}
for details. 
Furthermore, 
for the considered values of $q^2$,  the recoil-energy of the out-going
$K^*$ meson is large, and the seven
independent $B \to K^*$ form factors can be described in terms of
only two universal form factors \cite{Charles:1998dr}, 
which we denote as $\xi_\perp(q^2)$ and
$\xi_\parallel(q^2)$ for transversely and longitudinally polarized 
$K^*$ mesons, respectively
\cite{Beneke:2001wa}.

\FIGURE[hbt]{
      \psfig{file=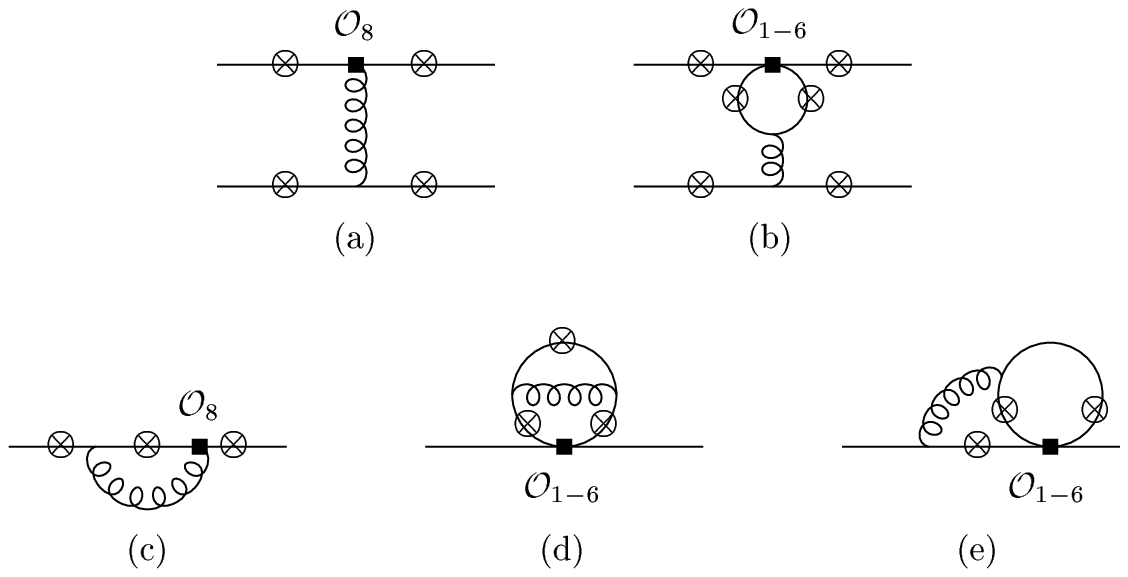, width=9cm}
\caption{Non-factorizable  NLO contributions to 
$\langle \gamma^*\bar K^*| H_{\rm eff} | \bar{B}\rangle$. 
Diagrams 
that follow from (c) and (e) by symmetry are not shown.}
\label{fig2}}

\FIGURE[t]{\epsfclipon
\epsfclipon\psfig{file=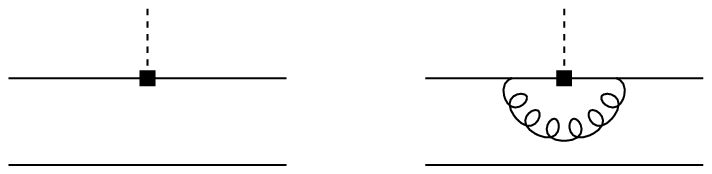, width=2.5cm }
       \hskip0.5cm
\epsfclipon       \psfig{file=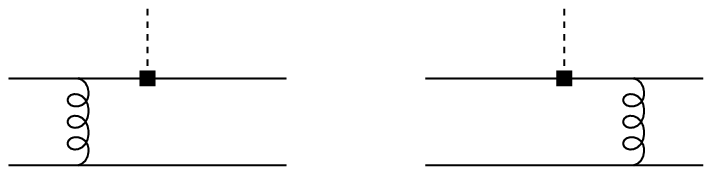, width=2.5cm}
       \hskip0.5cm
\epsfclipon       \psfig{file=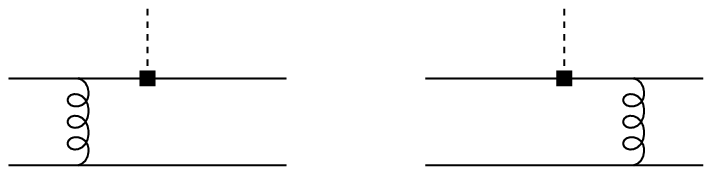, width=2.5cm }
\caption{Factorizable NLO corrections to the $B \to K^*$ form factors.}
\label{fig3}}

Factorizable next-to-leading order (NLO) form factor corrections 
are derived from Fig.~\ref{fig3} after
the corresponding infra-red divergent pieces are absorbed into
the {\em soft}\/ universal form factors  $\xi_\perp$ and $\xi_\parallel$,
see \cite{Beneke:2001wa} for details.
The non-factorizable vertex corrections (Fig.~\ref{fig2}c-e), are similar
to the NLO calculation for the {\em inclusive}\/ $b\to s\gamma^*$
transition, and the result for the two-loop
diagrams in Fig.~\ref{fig2}d+e are taken from
Ref.~\cite{Asatryan:2001de}. 
For the vertex corrections we chose
a renormalization scale $\mu = {\cal O}(m_b)$.
The non-factorizable 
hard-scattering corrections in Fig.~\ref{fig2}a+b and Fig.~\ref{fig1}c
involve the light-cone distribution amplitudes of both, $B$ and $K^*$ mesons.
(For  $q^2=0$ diagrams of this form have already been 
considered in \cite{Asatrian:1999mt}, but using
bound state model wave-functions, rather than light-cone distribution 
amplitudes.)
Since in these class of diagrams the typical quark- and gluon-virtuality 
is of order $\sqrt{\Lambda_{\rm QCD} m_b}$ we chose a different
renormalization scale $\mu'$ of that order.
In principle, we also have to consider NLO order corrections to the
annihilation graph in Fig.~\ref{fig1}c. However, since this term is
suppressed by  small Wilson coefficients $C_3$ and $C_4$ and
numerically small already at LO, we have neglected
these effects. Notice however, that the annihilation topology is 
numerically more important for $B \to \rho\gamma$ decays 
\cite{Grinstein:2000pc,Bosch:2001gv}.

The $B \to K^*\gamma$ decay rate is proportional
to the function $|{\cal T}_1(0)|^2 = |{\cal T}_2(0)|^2$.
In order to study the effect of NLO corrections it is
convenient to define a generalized exclusive ``Wilson'' coefficient
${\cal C}_7 \equiv {\cal T}_1(0)/\xi_\perp(0)$.
In Fig.~\ref{fig4} we have shown the $\mu$-dependence of  $|{\cal C}_7|^2$ 
at leading order (LO), including only next-to-leading order vertex
corrections (NLO$_1$), and including all next-to-leading order
corrections (NLO).
As expected, the NLO$_1$ vertex corrections cancel the 
renormalization-scale dependence of the LO result to a great extent.
(The hard-scattering corrections, arising at order $\alpha_s$ reintroduce
a mild scale-dependence.) 
Most importantly, we observe that the NLO corrections significantly
increase the theoretical prediction for  $|{\cal C}_7|^2$. 
Numerically,
we have  $|{\cal C}_7|^2_{\rm NLO} \simeq 1.78 \cdot |{\cal C}_7|^2_{\rm LO}$.
From this we predict the branching ratio as
\begin{equation}
\mbox{Br}(\bar{B}\to \bar K^*\gamma) = (7.9^{+1.8}_{-1.6})\cdot 10^{-5} 
\left(\frac{\tau_B}{1.6\mbox{ps}}\right) 
\left(\frac{m_{b,\rm PS}}{4.6\,\mbox{GeV}}\right)^{\!2}
\left(\frac{\xi_\perp(0)}{0.35}\right)^{\!2} 
\end{equation}
Comparing with the current experimental averages
\cite{Coan:1999kh} 
 $\mbox{Br}(\bar{B}^0\to \bar K^{*0}\gamma)_{\rm exp} = 
(4.54\pm 0.37)\cdot 10^{-5} $, 
 $\mbox{Br}(B^-\to \bar K^{*-}\gamma)_{\rm exp}= 
(3.81\pm 0.68)\cdot 10^{-5} $, 
and using the value $\xi_\perp(0) =0.35$ from QCD sum rules \cite{Ball:1998kk},
we observe that the central value of the
theoretical prediction overshoots the data by nearly a
factor of two. 
(An equivalent analysis with similar conclusions can be found
in Ref.~\cite{Bosch:2001gv}.)
\FIGURE[tbh]{
\psfig{file=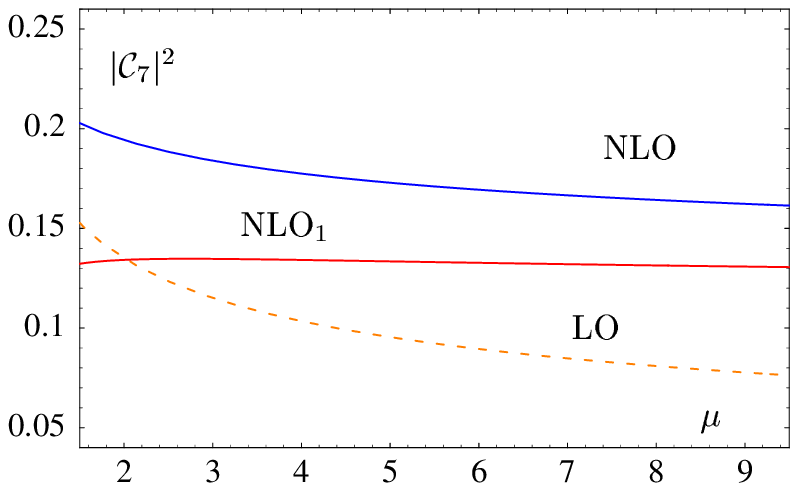, bb = 130 520 370 685, width=6.35cm }
\caption{
$|{\cal C}_7|^2$ as a function 
of the renormalization scale $\mu$, see text.}
\label{fig4}}
Possible explanations for this discrepancy are:
i) new physics contributions (this is rather unlikely because of the good
agreement between NLO theory and experiment for the {\em inclusive}
counterpart, $B \to X_s\gamma$), ii) sizeable $1/m_b$
power-corrections
(``chirally enhanced'' corrections play a role
for decays into light pseudoscalars \cite{Beneke:2001ev};
in our case, however, we expect a less dramatic effect),
iii) an insufficient understanding of the 
$B \to K^*$ form factors (a fit to the experimental data on the basis
of our formalism yields a somewhat smaller value,
$\xi_\perp(0) = 0.24 \pm 0.06$).

A quantity that is less sensitive to the precise value of $\xi_\perp(q^2)$
is provided by the $B \to K^*\ell^+\ell^-$
forward-backward asymmetry ${\cal A}_{\rm FB}$.
At LO the position of the asymmetry zero
$q_0^2$ is determined by the implicit relation
\begin{eqnarray}
 C_9 +  \mbox{Re}(Y(q_0^2))
 &=&  - \frac{2 M_B m_b}{q_0^2} \, C_7^{\rm eff} \ ,
\end{eqnarray}
and does not depend on form factors at all \cite{Burdman:1998mk}. 
As illustrated in Fig.~\ref{fig5} NLO corrections shift the 
position of the asymmetry zero from $q_0^2 = 3.4^{+0.6}_{-0.5}$~GeV$^2$ at LO
to $q_0^2 = 4.39^{+0.38}_{-0.35}$~GeV$^2$. (A slightly different value 
$q_0^2=3.94$~GeV$^2$ is found
if one takes the complete form factors from QCD sum rules \cite{Ball:1998kk},
instead of $\xi_\perp$ 
and the factorizable NLO corrections from \cite{Beneke:2001wa}).
In any case, a measurement of the forward-backward asymmetry zero provide
a clean test of the Wilson-coefficient $C_9$ in the standard model with
a rather small theoretical uncertainty of about 10\%.

\FIGURE[hbt]{
\psfig{file= 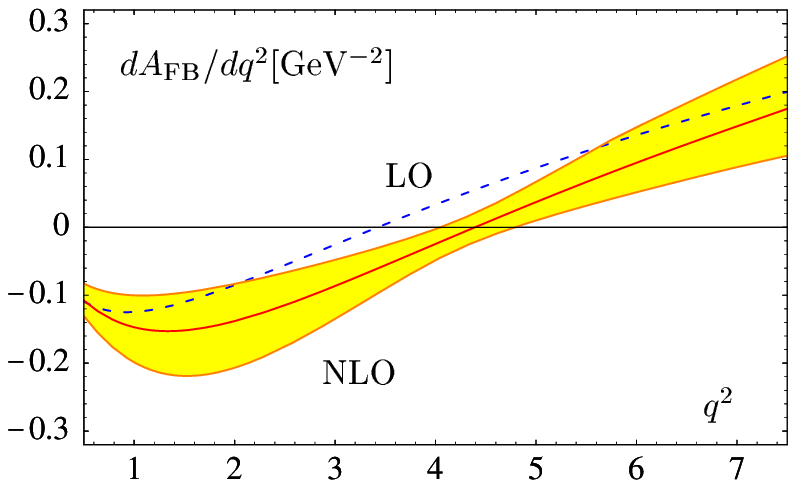, bb = 150 515  370 685,width=6cm }
\hskip1em
\psfig{file= 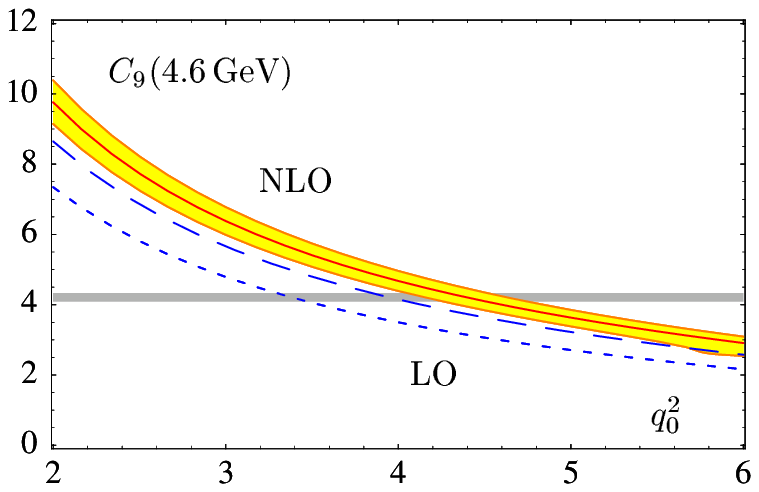, bb =  140 515  360 685, width=6cm }
\caption{The FB asymmetry as a function of $q^2$
  (left). The Wilson-coefficient $C_9$ as a function of
the FB asymmetry zero (right). 
The error band refers to a variation of all input parameters and
changing the renormalization scale between $m_b/2$ and $2 m_b$.
The dashed line is obtained from using the complete form factors
from \cite{Ball:1998kk}, see text. 
The grey band indicates the
standard model value. }
\label{fig5}
}

In summary, 
we have shown that a systematic improvement of the theoretical description
of exclusive radiative $B$ meson decays is possible. This is because in
the heavy quark limit decay amplitudes factorize into perturbatively
calculable hard-scattering kernels and universal 
soft form factors or light-cone distribution amplitudes, respectively.
The next-to-leading order corrections increase the branching ratio 
for the decay $B \to K^*\gamma$ by almost a factor of two (which is
at variance with the current experimental data if 
``standard'' values for the soft
form factors are used). They also shift the position of the forward-backward
asymmetry in the decay $B \to K^*\ell^+\ell^-$ towards $q_0^2 = 4.2 \pm 0.6$~GeV$^2$
in the standard model. In this case the precision of the prediction
is sufficient to test the Wilson coefficient $C_9$ with only 10\% theoretical
uncertainty.

\providecommand{\href}[2]{#2}\begingroup\raggedright\endgroup


\begin{thebibliography}{10}

\bibitem{Beneke:2001at}
M.~Beneke, T.~Feldmann, and D.~Seidel, 
\href{http://xxx.lanl.gov/abs/hep-ph/0106067}{{\tt hep-ph/0106067}} (to
  appear in Nucl.~Phys.~B).


\bibitem{Buchalla:1996vs}
G.~Buchalla, A.~J. Buras, and M.~E. Lautenbacher, 
{\em Rev. Mod. Phys.} {\bf 68} (1996) 1125--1144.


\bibitem{Beneke:1999br}
M.~Beneke, G.~Buchalla, M.~Neubert, and C.~T. Sachrajda, 
{\em Phys. Rev. Lett.} {\bf 83} (1999)
  1914, [\href{http://xxx.lanl.gov/abs/hep-ph/9905312}{{\tt hep-ph/9905312}}];
{\em Nucl. Phys.} {\bf
  B591} (2000) 313--418, [\href{http://xxx.lanl.gov/abs/hep-ph/0006124}{{\tt
  hep-ph/0006124}}].


\bibitem{Beneke:2001wa}
M.~Beneke and T.~Feldmann, 
{\em Nucl. Phys.} {\bf B592}
  (2000) 3--34, [\href{http://xxx.lanl.gov/abs/hep-ph/0008255}{{\tt
  hep-ph/0008255}}];\\
T.~Feldmann, 
{\em Nucl. Phys. Proc. Suppl.} {\bf
  93} (2001) 99--102, [\href{http://xxx.lanl.gov/abs/hep-ph/0008272}{{\tt
  hep-ph/0008272}}].


\bibitem{Charles:1998dr}
J.~Charles, A.~L. Yaouanc, L.~Oliver, O.~P{\`e}ne, and J.~C. Raynal, 
{\em Phys. Rev.} {\bf D60} (1999) 014001,
  [\href{http://xxx.lanl.gov/abs/hep-ph/9812358}{{\tt hep-ph/9812358}}].


\bibitem{Asatryan:2001de}
H.~H. Asatryan, H.~M. Asatrian, C.~Greub, and M.~Walker, 
{\em
  Phys. Lett.} {\bf B507} (2001) 162--172,  [\href{http://xxx.lanl.gov/abs/hep-ph/0103087}{{\tt hep-ph/0103087}}].


\bibitem{Asatrian:1999mt}
H.~H. Asatryan, H.~M. Asatrian, and D.~Wyler, 
{\em Phys. Lett.} {\bf B470} (1999) 223,
  [\href{http://xxx.lanl.gov/abs/hep-ph/9905412}{{\tt hep-ph/9905412}}].


\bibitem{Grinstein:2000pc}
A.~Ali and V.~M. Braun, 
{\em Phys.
  Lett.} {\bf B359} (1995) 223--235,
  [\href{http://xxx.lanl.gov/abs/hep-ph/9506248}{{\tt hep-ph/9506248}}];\\
%
B.~Grinstein and D.~Pirjol, 
{\em Phys. Rev.} {\bf D62} (2000) 093002,
  [\href{http://xxx.lanl.gov/abs/hep-ph/0002216}{{\tt hep-ph/0002216}}];\\
M.~Beyer, D.~Melikhov, N.~Nikitin, and B.~Stech, 
  \href{http://xxx.lanl.gov/abs/hep-ph/0106203}{{\tt hep-ph/0106203}}.


\bibitem{Bosch:2001gv}
S.~W. Bosch and G.~Buchalla, 
  \href{http://xxx.lanl.gov/abs/hep-ph/0106081}{{\tt hep-ph/0106081}}.


\bibitem{Coan:1999kh}
{\bf CLEO} Collaboration, T.~E. Coan {\em et.~al.}, 
{\em Phys. Rev. Lett.} {\bf 84} (2000)
  5283--5287, [\href{http://xxx.lanl.gov/abs/hep-ex/9912057}{{\tt
  hep-ex/9912057}}]; 
V.~Brigljevic [{\bf BaBar} Collaboration], at  
36th Rencontres de Moriond, 
March 2001, Les Arcs, France;
G.~Taylor [{\bf Belle} Collaboration], at  
36th Rencontres de Moriond, 
March 2001, Les Arcs, France.
 

\bibitem{Ball:1998kk}
P.~Ball and V.~M. Braun, 
{\em Phys. Rev.} {\bf D58} (1998) 094016,
  [\href{http://xxx.lanl.gov/abs/hep-ph/9805422}{{\tt hep-ph/9805422}}].


\bibitem{Beneke:2001ev}
M.~Beneke, G.~Buchalla, M.~Neubert, and C.~T. Sachrajda, 
  \href{http://xxx.lanl.gov/abs/hep-ph/0104110}{{\tt hep-ph/0104110}}.


\bibitem{Burdman:1998mk}
G.~Burdman, 
{\em Phys. Rev.} {\bf D57} (1998) 4254--4257,
  [\href{http://xxx.lanl.gov/abs/hep-ph/9710550}{{\tt
      hep-ph/9710550}}];\\
A.~Ali, P.~Ball, L.~T. Handoko, and G.~Hiller, 
 {\em Phys. Rev.} {\bf D61} (2000) 074024,
[\href{http://xxx.lanl.gov/abs/hep-ph/9910221}{{\tt
  hep-ph/9910221}}].

\end{thebibliography}
\end{document}